\documentclass[twocolumn,showpacs,preprintnumbers,amsmath,amssymb, nofootinbib]{revtex4}
\usepackage{graphicx}
\usepackage{dcolumn}
\usepackage{bm}

\begin{document}


\title{Asymptotic flatness at spatial infinity in higher dimensions}

\author{Kentaro Tanabe}
\affiliation{%
Yukawa Institute for Theoretical Physics, Kyoto University, Kyoto 606-8502, Japan
}%
\author{Norihiro Tanahashi}%
\author{Tetsuya Shiromizu}
\affiliation{
Department of Physics, Kyoto University, Kyoto 606-8502, Japan
}%

\date{\today}

\begin{abstract}
A definition of asymptotic flatness at spatial infinity in 
$d$ dimensions ($d\geq 4$) is given using the conformal completion 
approach. Then we discuss asymptotic symmetry and conserved quantities. 
As in four dimensions, in $d$ dimensions we should impose 
a condition at spatial infinity that the ``magnetic'' part 
of the $d$-dimensional Weyl tensor vanishes at faster rate than 
the ``electric'' part does, in order to realize the Poincare symmetry as asymptotic 
symmetry and construct the conserved angular momentum. 
However, we found that an additional condition should be imposed in 
$d>4$ dimensions.  
\end{abstract}

\maketitle

\section{Introduction}
If one considers an ``isolated'' system in general relativity, 
one should impose some asymptotic boundary conditions on 
gravitational fields. As one of such conditions, there is the 
asymptotically flat condition, which states that the metric 
should approach to Minkowski metric at ``far away'' place from 
gravitational sources. In order to define the notion of this ``far away'' 
covariantly, one often uses the conformal completion method 
introduced by Penrose~\cite{Penrose}. 
In this method, physical space-time $M$ is conformally embedded to 
unphysical space-time $\hat{M}$ with boundary, 
and this boundary is constituted of spatial infinity and null infinity. 
Hence, one can define asymptotic flatness, 
imposing some proper boundary conditions at this spatial infinity or null infinity. 

In four dimensions, asymptotic flatness at spatial infinity was investigated 
using the conformal completion method by Ashtekar and Hansen~\cite{AH}. 
They revealed that asymptotic symmetry at spatial infinity can be reduced 
to the Poincare symmetry which is a symmetry 
associated with ``background'' flat metric, and  constructed $4$-momentum 
and angular momentum. 
On the other hand, in higher dimensions,
there is only a few works about asymptotic structure at spatial 
infinity~\cite{Shiromizu:2004jt} or null infinity~\cite{Hollands:2003ie}
though recently the importance of higher dimensional black holes is 
increasing in string theory and TeV gravity scenario~\cite{HDBH,Review}.

While in four dimensions, uniqueness theorem was obtained~\cite{Israel}, 
we cannot prove the uniqueness for stationary 
black holes (counterexamples are Myers-Perry black hole~\cite{MP} and 
black ring~\cite{BR} with the same mass and angular momentum) 
in higher dimensions (although uniqueness was shown in~\cite{GIS} for 
static black holes). If one would like to classify these higher 
dimensional black holes using some parameters, the investigation on 
asymptotic structure at spatial infinity could play a key role. 

The purpose of this paper is to define asymptotic 
flatness and investigate asymptotic structure at spatial infinity in 
higher dimensions, following Ashtekar and Hansen~\cite{AH}.
(The reference~\cite{Shiromizu:2004jt} investigates into asymptotic 
flatness in higher dimensions following 
Ashtekar and Romano~\cite{Ashtekar:1991vb}. This analysis is useful 
when one is 
interested  only in spatial infinity. For full understanding of 
asymptotic structures, 
however, Ashtekar and Hansen's work is appropriate.)

The rest of this paper is organized as follows. In the 
section~\ref{sec:def}, we define asymptotic flatness at spatial 
infinity following Ashtekar and 
Hansen~\cite{AH}. In the section~\ref{sec:asym}, we investigate 
asymptotic structure: asymptotic symmetry and conserved quantities. 
Finally, we 
give a summary and discussion in the section~\ref{sec:summary}. 
In the appendix~\ref{App:DD} we introduce some important concepts 
in this literature such as 
directional dependence, and in the appendix~\ref{App:Mink}
we summarize basic features of conformal completion taking Minkowski 
space-time for an example.
Some important equations in this literature are derived in the 
appendix~\ref{App:derivations},
and in the appendix~\ref{App:d+1} we prove the equivalence of 
our expressions for conserved quantities  with the ADM formulae.

\section{Definition}
\label{sec:def}
We define asymptotic flatness at spatial infinity ($i^{0}$) in $d$ dimensions using the 
conformal completion method developed by Ashtekar and Hansen in four dimensions~\cite{AH}. 
In this paper, for simplicity we assume physical space-time $(M,g_{ab})$ 
satisfies the vacuum Einstein equation $R_{ab}=0$. It is easy to extend our current work to 
more general non-vacuum cases as long as one focuses on the asymptotically flat space-time. 

\textit{Definition}: $d$-dimensional physical space-time $(M,g_{ab})$ will be said to be 
asymptotically flat at spatial infinity~$i^{0}$ if there exists $(\hat{M},\hat{g}_{ab})$, where
$\hat{g}_{ab}$ is $C^{>d-4}$ at $i^{0}$ (see Appendix~\ref{App:DD} for the definition of 
$C^{>n}$), and embedding of $M$ into $\hat{M}$ satisfying
the following conditions:
\begin{enumerate}
\item
$\bar{J}(i^{0})=\hat{M}-M$, where $\bar{J}(i^0)$ is the closure of the union of 
chronological future and past of $i^0$.  
\item
There exists a function $\Omega$ on $\hat{M}$ that is $C^{2}$ at $i^{0}$ such that
$\hat{g}_{ab}=\Omega^{2}g_{ab}$ on $M$ and  
 $\hat{\nabla}_{a}\hat{\nabla}_{b}\Omega\hat{=}2\hat{g}_{ab}$,
$\Omega\hat{=}0$ and $\hat{\nabla}_{a}\Omega\hat{=}0$ at $i^{0}$ on $\hat{M}$.
\end{enumerate}
Here, 
and $\hat{\nabla}_{a}$ is  the connection for $\hat{g}_{ab}$, and
$\hat{=}$ implies the evaluation on $i^{0}$ (i.e.\ ``$=\lim_{\rightarrow i^0}$'' is equivalent to ``$\hat=$'').
The first condition requires that, in $\hat{M}$, $i^{0}$ is
connected to the points on $M$ only via spacelike curves. The second condition says that 
$\Omega$ behaves $\sim 1/r^2$ near $i^{0}$.  This is the same 
asymptotic behavior as in the Minkowski space-time (see Appendix~\ref{App:Mink}). 

Since we assume  $\hat{g}_{ab}$ is  $C^{>d-4}$ at $i^{0}$, 
$\hat{\partial}_{a_{1}}\cdots\hat{\partial}_{a_{(d-3)}}\hat{g}_{bc}$
has directional dependent limit at $i^{0}$ (where $\hat{\partial}$ is 
flat connection on $i^{0}$). This condition
is equivalent to  one such that $\Omega^{(5-d)/2}\hat{R}_{abcd}$ has 
directional dependent limit at $i^{0}$.
When we discuss asymptotic structure, we often use the Weyl tensor 
$\hat{C}_{abcd}$ as asymptotic gravitational fields. Thus,
it is convenient to use the latter condition on $\hat{R}_{abcd}$ for 
the discussions hereafter.

\section{asymptotic structure}
\label{sec:asym}
In this section, we show how to derive the asymptotic 
structure from the asymptotic flatness definition. 
Firstly, we discuss asymptotic symmetry in the section~\ref{Sec:symmetry}.
We show that the asymptotic symmetry is constituted of 
the Lorentz group and supertranslation group 
(infinite group of angular-dependent translation)
in higher dimensions. 
In the section~\ref{Sec:field}, we define asymptotic fields and study their transformation behavior 
under supertranslation.
We find  that supertranslation group reduces to the Poincare group
if we impose an additional 
asymptotic condition $B_{a_1a_2 \cdots a_{d-2}}\hat=0$ in the definition of  asymptotic 
flatness.
We define conserved quantities ($d$-momentum and angular momentum) associated to this Poincare symmetry 
in the section~\ref{Sec:energy}. We confirm that the conserved quantities we define
agree with the ADM formulae in this section and the appendix~\ref{App:d+1}.

\subsection{Asymptotic symmetry}
\label{Sec:symmetry}
The asymptotic symmetry is a group of mappings which conserve asymptotic structure. Here, by asymptotic structure we mean
$( \hat{g}_{ab}, \Omega^{(4-d)/2}\hat{\partial}\hat{g}_{bc})$ at $i^{0}$, since we impose $C^{>d-4}$ condition on the behavior of
$ \hat{g}_{ab}$ at $i^{0}$. 
In order to investigate this asymptotic symmetry, we consider the generator $\hat{\xi}$ of the asymptotic symmetry on $\hat{M}$. 
This generator $\hat{\xi}$ should be an extension of $\xi$, which is a generator of diffeomorphism on $M$. 
This extension $\hat{\xi}$ of $\xi$ to $i^{0}$ should satisfy
\begin{enumerate}
\item
$\hat{\xi}\hat{=}0$\ ,
\item
$\hat{\nabla}_{(a}\hat{\xi}_{b)}\hat{=}0$\ ,
\item
$\hat{\nabla}_{(a}\hat{\xi}_{b)}$ is a $C^{>d-4}$ tensor at $i^{0}$.
\end{enumerate}

Roughly speaking, these conditions set the behavior of components of $\hat{\xi}$ near $i^{0}$ as
\begin{equation}
\hat{\xi}^{a}\,\sim\,\frac{1}{r} + \frac{1}{r^{d-2}}.
\end{equation}
The first condition says that a generator $\hat{\xi}$ does not touch $i^{0}$. The second condition 
implies that $\hat{\xi}$ is asymptotically 
a  Killing vector, i.e.\ $\hat{g}_{ab}$ at $i^{0}$ is not changed.  
Before explaining the meaning of the third condition, let us consider the gauge freedom of the conformal completion. 
First, let $\omega$ be a function on $\hat{M}$, $C^{>d-4}$  at $i^{0}$ 
and $\omega\hat{=}1$. Then, a conformal 
completion such that $\hat{g}'_{ab} = (\omega\Omega)^2g_{ab}$ is equivalent to  $\hat{g}_{ab} = \Omega^2g_{ab}$, because 
$\omega\Omega$ satisfies
\begin{equation}
\omega\Omega\hat{=}0\, ,\, \hat{\nabla}_a(\omega\Omega)\hat{=}0\, , 
\, \hat{\nabla}_{a}\hat{\nabla}_{b}(\omega\Omega)
\hat{=}2\hat{g}_{ab} .
\end{equation}
Then, we cannot distinguish these two conformal completions under the asymptotic 
flatness definition in section~\ref{sec:def}.
This gauge freedom $\omega$ of the conformal completion reshuffles the 
value $\Omega^{(4-d)/2}\hat{\partial}\hat{g}_{bc}$ 
in the asymptotic structure as
\begin{align}
\Omega^{(4-d)/2}
\left(\hat{\nabla}_{a}'-\hat{\nabla}_{a}\right) \hat v_{b}
\quad & 
\notag \\
\;\hat=\;
\frac{1}{\omega}
\Bigl[\;\;
\delta^{c}_{a}\Omega^{(4-d)/2}\hat{\nabla}_{b}\omega 
&+\delta^{c}_{b}\Omega^{(4-d)/2}\hat{\nabla}_{a}\omega 
\notag \\
&
- g_{ab}\Omega^{(4-d)/2}\hat{\nabla}^{c}\omega
\;\;
\Bigr]
\hat v_{c}\;,
\end{align}
where $\hat{\nabla}_{a}'$ is the connection for $\hat{g}_{ab}'$ and $\hat v_a$ is any vector.
This equation can also be written as
\begin{equation}
\Omega^{(4-d)/2}\hat{\nabla}_{a}\hat{\nabla}_{(b}\hat{\xi}_{c)} 
\;\hat=\;
2\Omega^{(4-d)/2}(\hat{\nabla}_{a}\omega )\hat{g}_{bc}. 
\end{equation}
Thus, asymptotic structure $\Omega^{(4-d)/2}\hat{\partial}\hat{g}_{bc}$ 
has an ambiguity coming from gauge freedom $\omega$, 
and this ambiguity is reshuffled by order $1/r^{d-2}$ part of $\hat{\xi}$.  
Hence, asymptotic symmetry is the group of transformations
which does not change the asymptotic structure except for this gauge ambiguity.
Then,
we call this asymptotic symmetry transformation,
which is induced by order $1/r^{d-2}$ component of $\hat{\xi}$,
supertranslation group. 
As any two generators $\hat{\xi}^{1}$, $\hat{\xi}^{2}$ of 
supertranslation group commute:
\begin{align}
\left[\hat{\xi}^{1},\hat{\xi}^2\right]^{a} 
&\sim \frac{1}{r^{d-2}} \frac{\partial}{\partial U}r^{2-d} \sim \mathcal{O}\left(\frac{1}{r^{2d-5}}\right) \; ,
\end{align}
supertranslation group is abelian (where we use the fact that 
the contribution to 
$\Omega^{(4-d)/2}\hat\partial \hat g_{bc}$
from $\mathcal{O}(1/r^{2d-5})$ part of $\hat\xi$
is only
$\mathcal{O}(1/r^{d-3})$, which is regarded as zero at $i^0$,
and thus that part 
cannot transform the asymptotic structure). 
Because of angular dependence of $\omega$, however, supertranslation 
group has infinite translational directions. 
In this stage, asymptotic symmetry is not expected to be the Poincare symmetry.

\subsection{Asymptotic fields}
\label{Sec:field}
In order to construct conserved quantities associated with the 
asymptotic symmetry, 
we define asymptotic gravitational fields using the Weyl tensor 
$\hat C_{ambn}$ as\footnote{
In the definition of the magnetic part of the Weyl tensor~(\ref{eq:Mag}), 
the power 
of $\Omega$ is determined by the following evaluation.
Since $a_1, \cdots, a_{d-3}$ are indices for angular coordinates 
and $m$ is for the radial 
coordinate in polar coordinates, one of $p$ and $q$ has to be 
for the time coordinate $t$ and the 
other one has to be for an angular coordinate $\varphi$. 
Each parts in the magnetic part behaves near $i^0$ as 
$\hat \epsilon_{a_1\cdots a_{d-3}mpq}=\mathcal{O}(\sqrt{-\hat g})=\mathcal{O}(r^{2-d})$,
$\hat C^p{}_{qbn}=\mathcal{O}(r^{5-d})$,
$g_{tt}=\mathcal{O}(1)$,
$g_{\varphi\varphi}=\mathcal{O}(r^{-2})$ and $\hat\eta^a=\mathcal{O}(1)$.
Thus,
$\hat \epsilon _{a_{1}\cdots 
a_{d-3}mpq}\hat{C}^{pq}{}_{bn}\hat{\eta}^{m}\hat{\eta}^{n} 
=\mathcal{O}(r^{9-2d}) \sim \Omega^{(2d-9)/2}$, and we have to multiply an inverse of this factor to define 
a regular quantity.
\newline
$\hat E_{ab}$ is a symmetric traceless tensor since the Weyl tensor is traceless.
$\hat B_{a_1\cdots a_{d-3}b}$ is also a traceless tensor; $\hat B_{a_1\cdots a_{d-3}b}\hat g^{a_ia_j}=0$
due to antisymmetry of $\hat\epsilon$ in Eq.~(\ref{eq:Mag}); $\hat B_{a_1\cdots a_{d-3}b}\hat 
g^{a_i b}=0$ since it contains $\hat C^{[pqb]n}=0$.
This $\hat B_{a_1\cdots a_{d-3}b}$ is antisymmetric on the first $d-3$ indices $a_i$ ($i=1, \cdots, d-3$).
There are no symmetry between the last index $b$ and the other indices $a_i$ in general, though in 
the four-dimensional case the magnetic part $\hat B_{ab}$ is symmetric.
}
\begin{gather}
\hat E_{ab} \;\hat=\;  \Omega^{(5-d)/2}\hat{C}_{ambn}\hat{\eta}^{m}\hat{\eta}^{n} ,  \label{eq:Ele} \\
\hat B_{a_{1}\cdots a_{d-3}b} \;\hat=\;  \Omega^{(9-2d)/2}\hat \epsilon _{a_{1}\cdots
 a_{d-3}mpq}\hat{C}^{pq}{}_{bn}\hat{\eta}^{m}\hat{\eta}^{n} ,\label{eq:Mag}
\end{gather}
where $\hat \epsilon _{a_{1}\cdots a_{d-3}mpq}\equiv\sqrt{-\hat g}E_{a_{1}\cdots a_{d-3}mpq}$ is a totally antisymmetric tensor in $\hat M$,
and we take the convention that $E_{012\cdots d-1}=1$.
$\hat{\eta}_{a}\hat=\hat{\nabla}_{a}\Omega^{1/2}$ is a 
normal vector to $\Omega=\text{constant}$ surface which becomes a unit vector at $i^0$. 
We call these asymptotic fields~(\ref{eq:Ele}) and (\ref{eq:Mag}) electric and magnetic parts of 
the Weyl tensor respectively. 
As these fields do not 
have components parallel to $\hat{\eta}^{a}$, we can regard them
 as fields on a timelike hypersurface $\mathcal{S}$ normal to $\hat \eta^a$.

Firstly, let us derive asymptotic field equations.
Using the Bianchi identity in the physical vacuum space-time $\nabla_{[m}C_{ab]cd}=0$, we obtain the following 
equation in terms of the unphysical space-time quantities:
\begin{equation}
\hat{\nabla}_{[m}\hat{C}_{ab]cd} = \Omega^{-1}\left( \hat{g}_{c[m}\hat{C}_{ab]pd}\hat{\nabla}^{p}\Omega + \hat{g}_{d[m}\hat C_{ab]cp}
\hat{\nabla}^{p}\Omega\right)\;.  \label{eq:a}
\end{equation}
It is better to rewrite the left-hand side as
\begin{align}
\hat{\nabla}_{[m}\hat{C}_{ab]cd} = \Omega^{-1}
\biggl[
\Omega^{(d-3)/2}\hat{\nabla}_{[m}
\!\Bigl(
\Omega^{(5-d)/2}\hat{C}_{ab]cd}\Bigr)
\;\;
& \notag \\
-\frac{5-d}{2}
(\hat{\nabla}_{[m}\Omega ) \hat{C}_{ab]cd}  
&\biggr],
\label{eq:a2}
\end{align}
since
$\Omega^{(5-d)/2}\hat{C}_{abcd}$ have directional dependent limit at 
$i^{0}$.  We project these equations into the timelike hypersurface $\mathcal{S}$, 
and contract with $\hat \eta^{a}$. Then, we get the equations for the electric part
\begin{equation}
\hat D_{a}\hat E_{bc}-\hat D_{b}\hat E_{ac}  
 \;\hat=\; (4-d)\hat{h}_{a}^{~p}\hat{h}_{b}^{~q}\hat{h}_{c}^{~r} \Omega^{(5-d)/2}\hat{C}_{pqrm}\hat{\eta}^{m} 
\label{eq:f}
\end{equation}
and for the magnetic part
\begin{align}
\hat D_{b}\hat B_{a_{1}\cdots a_{d-3}c}-\hat D_{c}&\hat B_{a_{1}\cdots a_{d-3}b}   \label{eq:Beq} \\ 
\hat = -(d-3)&\Omega^{(9-2d)/2}
\hat \epsilon_{a_{1}\cdots a_{d-3}}{}^{fpq} \,^{(d-1)}\hat{C}_{bcpq}\hat{\eta}_{f}\, , \notag
\end{align}
where $\hat h_{ab}$ is the induced metric on $\mathcal{S}$, and 
\begin{equation}
 \hat D_{a}\hat  v_{b}\equiv\Omega^{1/2}\hat h_{a}^{~p}\hat h_{b}^{~q}\hat{\nabla}_{p}\hat v_{q}
\end{equation}
is a regular differentiation with respect to $\hat{h}_{ab}$ on $\mathcal{S}$. $^{(d-1)}\hat{C}_{abcd}$ is 
the $(d-1)$-dimensional Weyl tensor with respect to $\hat{h}_{ab}$, and $\Omega^{(5-d)/2}\,^{(d-1)}\hat{C}^a{}_{bcd}$ 
have a directional dependent limit at $i^{0}$. 
(For detailed derivations of Eqs.~(\ref{eq:f}) and (\ref{eq:Beq}), see Appendix~\ref{App:firstset}.)

Next, in order to see how these fields transform under the supertranslation, 
we introduce potentials of the Weyl tensor. To do so, we will use 
the Bianchi identity in the unphysical space-time
\begin{equation}
 \hat{\nabla}_{m}\hat{C}_{abc}{}^{m} +\frac{2(d-3)}{d-2}\hat{\nabla}_{[a}\hat{S}_{b]c}=0 \;\; ,
\label{eq:b}
\end{equation}
where
\begin{equation}
\hat{S}_{ab}\equiv\hat{R}_{ab}-\frac{\hat{R}}{2(d-1)}\hat{g}_{ab} \;\;.
\label{eq:S}
\end{equation}
Since we assume $\hat{g}_{ab}$ to be $C^{>d-4}$, $\Omega^{(5-d)/2}\hat{S}_{ab}$ admits directional dependent limit at $i^{0}$.
Then, we define potentials as
\begin{align}
\hat E\;&\hat=\;
\Omega^{(5-d)/2}\hat{S}_{pq} \hat{\eta}^{p}\hat{\eta}^{q}\;\;\;,\\
\hat Q_{a}\;&\hat=\;
\Omega^{(5-d)/2}\hat{S}_{pq}\hat{h}_{a}^{~p}\hat{\eta}^{q} \;\;,\\
\hat U_{ab}\;&\hat=\;
\Omega^{(5-d)/2}\hat{S}_{pq}\hat{h}_{a}^{~p}\hat{h}_{b}^{~q} \;.
\end{align}
Using Eqs.~(\ref{eq:a}) and (\ref{eq:b}), we can write down 
the electric  and the magnetic part 
in terms of potentials as
\begin{equation}
 \hat E_{ab}
\,\hat=\,
\frac{-1}{2(d-2)}
\left[
\frac{1}{d-3}\hat D_{a}\hat D_{b}\hat E+\hat E \hat{h}_{ab}
+(4-d)\hat U_{ab}
\right],
\label{elepo}
\end{equation}
\begin{align}
\!
\hat B_{a_{1}\cdots a_{d-3}b}
\,\hat=\,
\frac{-1}{d-2}&\hat \epsilon _{a_{1}
\cdots a_{d-3}mpq}\hat\eta^m\Omega^{(4-d)/2}
\notag \\
&\qquad\times
\hat D^{p}\left(
\hat U^{q}_{~b}
 -\frac{1}{d-3}\hat E \hat{h}^{q}_{~b}
\right)
\notag \\
\equiv \frac{-1}{d-2}
&\hat \epsilon _{a_{1}\cdots a_{d-3}mpq}
\Omega^{(4-d)/2} \hat\eta^m\hat D^{p}\hat{\mathcal{K}}^{q}_{~b}\,\, ,
\label{eq:mag2}
\end{align}
where we define a tensor $\hat{\mathcal{K}}_{ab}$ by Eq.~(\ref{eq:mag2}). (Eqs.~(\ref{elepo}) 
and (\ref{eq:mag2}) are derived in Appendix~\ref{App:secondset}.) 

Now, we observe transformation behaviors of the asymptotic fields under the 
supertranslation. 
In a supertranslational transformation 
$\hat{g}_{ab} \rightarrow \hat{g}_{ab}'=\omega^{2}\hat{g}_{ab}$,
where $\omega$ is a $C^{>d-4}$ function ($\omega\hat{=}1$), $\hat{S}_{ab}$ transforms as
\begin{align}
\hat{S}_{ab}' = \hat{S}_{ab}&- (d-2)\omega^{-1}\hat{\nabla}_{a}\hat{\nabla}_{b}\omega  \nonumber \\ 
& +2(d-2)\omega^{-2}
(\hat{\nabla}_{a}\omega)(\hat{\nabla}_{b}\omega)  \nonumber \\ 
&+\frac{2-d}{2}\omega^{-2}\hat{g}_{ab}(\hat{\nabla}_{m}\omega)(\hat{\nabla}^{m}\omega)\; .\label{conf}
\end{align}
Since $\omega$ is $C^{>d-4}$ and $\omega\hat=1$, it can be written as
\begin{equation}
 \omega=1+\Omega^{(d-3)/2}\alpha\;,
\end{equation}
where $\alpha$ is a function which has directional dependent limit 
at $i^{0}$. 
Then, 
the potentials $\hat E$ and $\hat U_{ab}$ transform under the supertranslational transformation as 
\begin{gather}
\hat E'\;\hat=\;\hat E-(d-2)(d-3)(d-4)\alpha\;, \label{E}\\
\hat U_{ab}'\;\hat=\;\hat U_{ab} -(d-2)\left( \hat D_{a}\hat D_{b}\alpha +(d-3)\alpha \hat{h}_{ab}\right)\;.\label{U}
\end{gather}
To show these equations, we use a relation
\begin{align}
\Omega^{(4-d)/2}\hat \eta^a \hat \nabla_a \omega
&\;\hat =\;\Omega^{1/2}\hat \eta^a \hat \nabla_a 
\alpha+(d-3)\alpha \notag \\
&\;\hat =\;(d-3)\alpha\; .
\end{align}
The second equality in this relation holds since $\alpha$ has directional dependent limit at $i^0$ and 
$\hat\eta^a\hat\nabla_a \alpha\hat=0$.
We note that only $\hat\nabla_a\hat\nabla_b\omega$ term of Eq.~(\ref{conf}) contributes to the 
variation of $\hat E$ and $\hat U_{ab}$.

It is easy to check that the electric part does not change in this transformation. 
On the other hand, the potential of the magnetic part $\hat{\mathcal{K}}_{ab}$ transforms as
\begin{equation}
\hat{\mathcal{K}}_{ab}'\;\hat=\;\hat{\mathcal{K}}_{ab} -(d-2)(\hat D_{a}\hat D_{b}\alpha +\alpha \hat{h}_{ab})\;.
\label{K-trans}
\end{equation}
Hence, the magnetic part $\hat B_{a_{1}\cdots a_{d-3}b} $ does change under the supertranslational transformation. 

\subsection{Conserved quantities and Poincare symmetry}
\label{Sec:energy}
Let us construct conserved quantities and the asymptotic symmetry in this section. 
First, as in four dimensions, we impose an additional 
condition
\begin{equation}
\hat B_{a_{1}\cdots a_{d-3}b} \;\hat=\;0\;.\label{eq:c}
\end{equation}
This condition implies that the Taub-NUT charge is zero. 
Although it is of course possible to consider asymptotically locally 
Minkowski space-time with $\hat B_{a_{1}\cdots a_{d-3}b} \neq 0$, 
we focus only on asymptotically globally Minkowski space-time 
in this paper. 
In order to impose the condition~(\ref{eq:c}) consistently with Eq.~(\ref{eq:Beq}), 
we must require a further additional condition 
\begin{equation}
\Omega^{(5-d)/2}\,^{(d-1)}\hat{C}^a{}_{bcd} \;\hat=\;0 \label{eq:d}
\end{equation}
as one of the conditions in the definition of asymptotic flatness.
Note that 
$^{(d-1)}C_{abcd}$ vanishes automatically in four dimensions. 
By the way, the condition~(\ref{eq:c}) 
is not preserved under the supertranslation. To preserve the condition~(\ref{eq:c}), we realise 
that one must impose 
\begin{equation}
\hat D_{a}\hat D_{b}\alpha +\alpha \hat{h}_{ab}\;\hat=\;0\;. \label{eq:e}
\end{equation}
As in four dimensions, we can write down the solution to 
Eq.~(\ref{eq:e}) as $\alpha=\hat\omega_{a}\hat\eta^{a}$, where $\hat\omega_a$ 
is a fixed vector at $i^{0}$. The number of independent solutions 
is the number of dimensions. Thus, we can regard the transformation 
generated by $\alpha$ satisfying Eq.~(\ref{eq:e}) as translation. 
Then, the asymptotic symmetry reduces to the Poincare 
group which is constituted of the Lorentz group and the translation group,
and we can define conserved quantities associated with this  Poincare 
symmetry. 

Now, it is ready to define conserved quantities. 
First, we define $d$-momentum $P_{a}$ for translation $\hat \omega^{a}$ as
\begin{equation}
P_{a}\omega^{a} \equiv \frac{-1}{8\pi G_{d}(d-3)} \int_{S^{d-2}} 
\!\!\!\!\!\!\!
\hat E_{ab}\hat \omega^{a}
\hat \epsilon ^{b}{}_{e_{1}\cdots e_{d-2}m}
\hat\eta^m
dS^{e_{1}\cdots e_{d-2}},
\label{eq:g}
\end{equation}
where $dS^{e_{1}\cdots e_{d-2}}$ is the volume element on $(d-2)$-dimensional 
unit sphere $S^{d-2}$ on $i^{0}$. From Eq.~(\ref{eq:f}), 
we get $\hat D_{a}\hat E^{ab}\hat=0$ since $\hat E_{ab}$ is traceless. 
Then, the integral 
of Eq.~(\ref{eq:g}) is independent of the choice of time slice at 
$i^{0}$, and thus $P_a\omega^a$ is conserved. 
After tedious calculations, we can show that Eq.~(\ref{eq:g}) agrees  
with the ADM formula (see Appendix~\ref{App:energy} and \ref{App:momentum}). 

Next, in order to define angular momentum using the magnetic part of the Weyl tensor, we consider 
the next-to-leading order part of $\hat B_{a_{1}\cdots a_{d-3}b}\,$:
\begin{equation}
\hat \beta_{a_{1}\cdots a_{d-3}b} 
\;\hat=\; 
\Omega^{4-d}\hat \epsilon _{a_{1}\cdots
 a_{d-3}mpq}\hat{C}^{pq}{}_{bn}\hat{\eta}^{m}\hat{\eta}^{n} .
\end{equation}
Since $\hat \beta_{a_{1}\cdots a_{d-3}b}$ satisfies $\hat D_{b}\hat \beta^{ba_{2}\cdots a_{d-3}c}\hat=0$
due to Eq.~(\ref{eq:Beq}) and the traceless property of $\hat B_{a_1\cdots a_{d-3}b}$, 
we can define conserved quantity 
$M_{ab}$ which is regarded as angular momentum: 
\begin{align}
M_{ab}F^{ab}
\equiv
\frac{-1}{8\pi G_{d}(d-2)!} 
\int_{S^{d-2}}
\hat \beta_{a_{1}\cdots a_{d-3}b}
&\xi^{a_{1}\cdots a_{d-3}}
\;
\label{eq:h}
\\
\times \hat \epsilon ^{b}_{~e_{1}\cdots e_{d-2}m}
&\hat\eta^m dS^{e_{1}\cdots e_{d-2}}, 
\notag
\end{align}
where
\begin{equation}
\xi^{a_{1}\cdots a_{d-3}} \equiv \hat \epsilon ^{a_{1}\cdots a_{d-3}mpq}\hat{\eta}_{m}F_{pq}
\end{equation}
and $F_{ab}$ is any skew tensor in $\cal S$. 
The coefficients in (\ref{eq:h}) so 
that angular momentum $M_{ab}$ transforms  properly under translation 
$\hat \omega_{a}$, including the coefficient:
\begin{equation}
M_{ab} \rightarrow M_{ab}'=M_{ab} + 2P_{[a}\hat \omega_{b]}\;. \label{eq:i}
\end{equation} 
 See Appendix~\ref{App:angular} for details of the coefficient determination.

\section{Summary and discussion}
\label{sec:summary}
In this paper, we gave a definition of asymptotic flatness, and constructed 
conserved quantities, $d$-momentum and angular momentum in $d$ dimensions. 
As in four dimensions, by imposing an additional constraints on the behavior 
of the ``magnetic'' part of the Weyl tensor, we can remove 
the supertranslational ambiguity. Then, the asymptotic symmetry of the space-time 
reduces to the Poincare symmetry, which is a symmetry of ``background''  
flat metric, and we can construct conserved quantities associated with this 
Poincare symmetry. It can be shown that 
the expressions of these conserved quantities agree with the ADM formulae.  

In four dimensions, the additional constraint is only $\hat{B}_{ab}=0$ to 
realize the Poincare symmetry as the asymptotic symmetry,
and it is satisfied if there is a Killing vector in $M$,
such as timelike Killing vector $(\partial / \partial t)$ or rotational 
Killing vector $(\partial /\partial \varphi )$~\cite{AM2}. 
On the other hand, in higher dimensions, due to the evolution equation~(\ref{eq:Beq})  
of $\hat B_{a_{1}\cdots a_{d-3}b}$, we need to impose a further condition 
$\Omega^{(5-d)/2}\,^{(d-1)}\hat C^a{}_{bcd}\hat{=}0$ to remove the supertranslational 
ambiguity and realize the Poincare symmetry. As in four 
dimensions, $\hat  B_{a_{1}\cdots a_{d-3}b}=0$ 
would be satisfied in stationary or axisymmetric space-time in higher 
dimensions. However, it might be interesting to investigate asymptotic symmetry 
under more general conditions which 
$\Omega^{(5-d)/2}\,^{(d-1)}\hat C^a{}_{bcd}\hat{=}0$ does not hold. 

In this paper, we focused only on spatial infinity. However, it is interesting to 
explore the full asymptotic structure including null infinity. 
As our future work, we would investigate the relationship between the Bondi 
energy formula at null infinity and Weyl tensor formula in this paper at 
spatial infinity. 
We also would like to consider asymptotic structure and its symmetry at null 
infinity, and investigate its connection to the supertranslation at spatial 
infinity. 

Another future issue is the preparation for the uniqueness theorem in 
stationary black hole space-times. As mentioned in the introduction, at first 
glance, the uniqueness theorem does not hold in higher dimensions, although there are some 
partial achievements~\cite{Hollands,Morisawa,Hollands2,Morisawa2}. However, we would guess that 
the reason why we fail to prove it is due to lack of asymptotic boundary conditions. 
If we can specify the boundary condition appropriately, we will be able to 
prove the uniqueness theorem. The mass, charge and angular momentum are not 
enough to specify the black hole space-time uniquely. The additional information 
for the uniqueness may be higher multipole moments. Therefore, the study on 
higher multipole moments in stationary space-time will be useful. 

\begin{acknowledgments}
The work of TS was supported by Grant-in-Aid for Scientific Research from Ministry 
of Education, Science, Sports and Culture of Japan (Nos.~19GS0219 and 20540258). NT was supported by JSPS
Grant-in-Aid for Scientific Research No.~20$\cdot$56381. 
This work was supported by the Grant-in-Aid for the Global COE Program 
"The Next Generation of Physics, Spun from Universality and Emergence" from the Ministry of Education, 
Culture, Sports, Science and Technology (MEXT) of Japan.
\end{acknowledgments}

\appendix

\section{directional dependence}
\label{App:DD}
In the conformal completion method, spatial infinity which has a non-zero size in the physical space-time~$M$ contracts to a point~$i^{0}$ 
in the unphysical space-time~$\hat{M}$. Hence, the definition of differentiability and continuity of physical fields (e.g.\ electromagnetic 
fields or gravitational fields) on $i^{0}$ is more subtle. In this appendix, 
we give the notion of directional dependent limit and $C^{>n}$ class.  

First, the tensor $\hat{T}^{a\cdots b}_{c\cdots d}$ is said to have directional dependent limit at $i^{0}$ if $\hat{T}^{a\cdots b}_{c\cdots d}$ satisfies 
the following conditions:
\begin{enumerate}
\item \hfill
$\displaystyle
\lim_{\rightarrow i^{0}} \hat{T}^{a\cdots b}_{c\cdots d} =\hat{T}^{a\cdots 
 b}_{c\cdots d}(\hat{\eta}) \;, 
$\hfill\mbox{}
\hfill
\newline
where $\hat{\eta}$ is a vector on tangential space at $i^{0}$, which is tangent to the curve arriving at $i^{0}$. 
\item
The derivative coefficients at $i^{0}$ defined by
\begin{equation}
\left(\Omega^{1/2}\hat{\nabla}_{e_{1}}\right)\cdots\left(\Omega^{1/2}\hat{\nabla}_{e_{n}}\right)\hat{T}^{a\cdots b}_{c\cdots d} \notag
\end{equation} 
are regular. 
\end{enumerate}
The first condition says that, since $i^{0}$ has a non-zero size ($S^{d-2}$) in 
$M$, $\hat{T}^{a\cdots b}_{c\cdots d}$ may have an angular 
dependence even in the limit $r\rightarrow \infty$. 
The operator $\Omega^{1/2}\hat \nabla_a$ in the second 
condition gives regular derivative coefficients,
since an application of a derivative operator 
$\hat \nabla_a$ in $\hat{M}$  corresponds to a multiplication of 
$r$ near $i^0$ (see Appendix~\ref{App:Mink}).
The second condition says that these regular derivative coefficients should be finite and regular.

Next, we define $C^{>n}$ class. A tensor $\hat{T}^{a\cdots b}_{c\cdots d}$ is $C^{>n}$ at $i^{0}$ if 
the $n+1$ derivatives of 
$\hat{T}^{a\cdots b}_{c\cdots d}$ have directional dependent limit at $i^{0}$. 
For example, when we set $\hat{g}_{ab}$ to be $C^{>n}$ at $i^{0}$, the behavior of $\hat{g}_{ab}$ near $i^{0}$ is 
\begin{equation}
\hat{g}_{ab} \sim \text{const.} + \frac{f(\theta,\varphi,\cdots)}{r^{n+1}} \; ,
\label{eq:metric}
\end{equation} 
where the dots stand for other angular coordinates.

\section{conformal completion for Minkowski space-time}
\label{App:Mink}
In this appendix, we discuss conformal completion for Minkowski space-time. This analysis 
tells us how we can define asymptotically flat space-time in general. 
First, we introduce coordinates $(U,V)$ such that
\begin{align}
ds^2=&-dt^{2}+dr^{2} +r^{2}d\Omega_{d-2}^2 \nonumber \\
=&-du dv +\frac{(u-v)^2}{4}d\Omega_{d-2}^2 \nonumber \\ 
=&-\frac{dUdV}{\cos ^2U\cos^2V} + \frac{\sin^2(U-V)}{4\cos^2 U\cos^2 V}d\Omega_{d-2}^{2} \;,
\label{eq:metricMink}
\end{align}
where
\begin{equation}
u=t-r=\tan U~,~v=t+r=\tan V \;,
\end{equation}
and $d\Omega_{d-2}^{2}$ is a metric on unit $S^{d-2}$. 
Let us take $\Omega\equiv\cos U\cos V$ as a conformal factor. In this case, we can see that
\begin{equation}
\hat{\nabla}_{a}\hat{\nabla}_{b}\Omega \;\hat{=}\; 2\hat{g}_{ab}
\end{equation}
holds at $i^0$. The unit normal vector $\hat{\eta}_{a}$ to 
$\Omega=\text{constant}$ surface becomes
\begin{equation}
\hat{\eta}_{a} \hat{=}  \hat{\nabla}_{a}\Omega^{1/2}
\end{equation}
on $i^{0}$. 

It will be useful for  discussions in the main text to look how the differential operators behave:
\begin{equation}
\hat{\nabla}_{U} \sim \frac{\partial}{\partial U} \sim r^2\frac{\partial}{\partial r}\; ,
\end{equation}
i.e.\ an application of $\hat\nabla_a$ corresponds to a multiplication of $r$.
When we say that $\hat{g}_{ab}$ is $C^{>n}$ at $i^{0}$, by the way, we should take differentiation 
in the coordinates $(U,V)$, and so this condition implies
that the metric in the unphysical space-time 
is given by
\begin{equation}
\hat{g}_{ab} =\hat{\eta}_{ab}\left( 1 + \frac{f(\theta ,\varphi ,\cdots)}{r^{n+1}}\right),
\end{equation} 
where 
\begin{equation}
 \hat\eta_{ab}dx^adx^b\equiv -dUdV + \frac{\sin^2\left(U-V\right)}{4}d\Omega^2_{d-2}
\end{equation}
is the unphysical space-time metric corresponding to the flat metric~(\ref{eq:metricMink}) in the physical space-time. 

\section{Derivations of Eqs.~(\ref{eq:f}), (\ref{eq:Beq}), (\ref{elepo}) and (\ref{eq:mag2})}
\label{App:derivations}
In this appendix, we give detailed derivations of Eqs.~(\ref{eq:f}), 
(\ref{eq:Beq}), (\ref{elepo}) and (\ref{eq:mag2}). 
Since we compute quantities only at spatial infinity,
we omit ``hat'' and $\lim_{ \rightarrow i^0}$ 
throughout this appendix
for convenience.

\subsection{Derivation of Eqs.~(\ref{eq:f}) and (\ref{eq:Beq})}
\label{App:firstset}
First, from Eqs.~(\ref{eq:a}) and (\ref{eq:a2}), 
we obtain
\begin{align}
&\Omega^{1/2}\nabla_{[m}\mathcal{X}_{ab]cd} 
\nonumber \\
&\;\;
= \Omega^{-1/2}\Bigl( g_{c[m}\mathcal{X}_{ab]pd}\nabla^{p}\Omega +g_{d[m}\mathcal{X}_{ab]cp}\nabla^{p}\Omega \notag \\
&\qquad\qquad\qquad
\qquad\qquad\quad
+\frac{5-d}{2}(\nabla_{[m}\Omega)\mathcal{X}_{ab]cd} 
\Bigr),
\label{eleom}
\end{align}
where $\mathcal{X}_{abcd}\equiv\Omega^{(5-d)/2}C_{abcd}$ . 

Multiplying $\eta^b \eta^d h_e^m h_f^a h_g^c$ to the above, the left-hand side 
becomes 
\begin{align}
&\Omega^{1/2} \eta^b \eta^d h_e^m h_f^a h_g^c \nabla_{[m}\mathcal{X}_{ab]cd}
\nonumber \\
&\;\;\;\;
=\frac{1}{3}(D_e E_{fg}-D_f E_{eg}) 
-\frac{1}{3}(2W_{feg}+W_{gef}-W_{gfe}),
\end{align}
where we used the fact that
\begin{equation}
\Omega^{1/2}\nabla_{a}\eta_{b} = g_{ab} - \eta_{a}\eta_{b} = h_{ab} \; ,
\end{equation}
and the definition 
\begin{equation}
W_{abc}\equiv h_a^e h_b^f h_c^g \mathcal{X}_{efgd}\eta^d. 
\end{equation}
In addition, we used the fact that $\mathcal{X}_{abcd}$ has directional dependent limit 
and thus $\eta^e \nabla_e \mathcal{X}_{abcd}$ vanishes. 
In the right-hand side of Eq.~(\ref{eleom}),
the second and third terms become 
\begin{equation}
\Omega^{-1/2}\eta^b \eta^d h_e^m h_f^a h_g^c
g_{d[m}\mathcal{X}_{ab]cp}\nabla^p \Omega = \frac{2}{3}W_{efg}
\end{equation}
and
\begin{equation}
\frac{5-d}{2}\Omega^{-1/2}\eta^b \eta^d h_e^m h_f^a h_g^c
\nabla_{[m}\Omega \mathcal{X}_{ab]cd}=
\frac{5-d}{3}W_{efg}.
\end{equation}
The first term vanishes since $\nabla^p\Omega=2\Omega^{1/2}\eta^p$. 
Finally, we obtain Eq.~(\ref{eq:f}) 
from Eq.~(\ref{eleom}), that is
\begin{align}
D_e E_{fg}-D_fE_{eg} & =  (d-5)W_{feg}+W_{gef}+W_{fge}\nonumber \\
& =  (d-4)W_{feg}\;,
\end{align}
where we used $W_{[abc]}=0$ in the second line. 

Next, we multiply 
$\Omega^{(4-d)/2}
\mathcal{E}_{a_1\cdots a_{d-3}}{}^{fcd}
\eta_f \eta^b h_g^m h_h^a$ 
to Eq.~(\ref{eleom}),
where 
$\mathcal{E}_{a_1\cdots a_{d-3}}{}^{fcd}\equiv
h_{a_1}^{b_1} \cdots h_{a_{d-3}}^{b_{d-3}}\epsilon_{b_1 \cdots b_{d-3}}{}^{fcd}$,
 and then obtain 
\begin{align}
&\frac{1}{3}\left(
D_g B_{a_1 \cdots a_{d-3}h}-D_h B_{a_1 \cdots a_{d-3}g} 
\right) \nonumber \\
&-\frac{1}{3}
{\cal E}_{a_1 \cdots a_{d-3}}{}^{fcd} \Omega^{(4-d)/2}
\nonumber \\
&\;\;\;\;
\times\left(
h_{fg}h^a_h \eta^b+h_g^b h_h^a\eta_f -h_{hf}h_g^a \eta^b-h_h^b h_g^a \eta_f 
\right)
\mathcal{X}_{abcd} 
\nonumber \\
& =\frac{1}{3}\Bigl[ 4\left(h_{dg}E_{hc}-h_{dh}E_{gc}\right)+ 
(5-d)h_g^a h_h^b \mathcal{X}_{abcd} \Bigr] 
\nonumber \\
& ~~~\times 
\Omega^{(4-d)/2}{\cal E}_{a_1 \cdots a_{d-3}}{}^{fcd}\eta_f\; .
\label{eq:Beq2}
\end{align}
From this equation, we obtain Eq.~(\ref{eq:Beq}):
\begin{align}
&\!\!
D_g B_{a_1 \cdots a_{d-3}h}-D_h B_{a_1 \cdots a_{d-3}g} \nonumber \\
& =-(d-3)\Omega^{(9-2d)/2}h_g^p h_h^q \eta_f \mathcal{E}_{a_1 \cdots a_{d-3}}^{~~~~~~~~~fcd}
\,^{(d-1)}C_{pqcd}\;,
\label{eq:Beq3}
\end{align}
where $^{(d-1)}C_{abcd}$ is the $(d-1)$-dimensional Weyl tensor 
on $\Omega=\text{constant}$ surface at $i^0$. To transform Eq.~(\ref{eq:Beq2}) to Eq.~(\ref{eq:Beq3}),
we used the following relations
\begin{align}
&{\cal E}_{a_1 \cdots a_{d-3}}{}^{fcd} h_{hf}h_g^a \eta^b \mathcal{X}_{abcd} \nonumber \\
&~={\cal E}_{a_1 \cdots a_{d-3}}{}^{fcd} h_{hf}h_g^a \eta^b (h_c^i+\eta_c \eta^i)
(h_d^j+\eta_d \eta^j) \mathcal{X}_{abij} \nonumber \\
&~=2{\cal E}_{a_1 \cdots a_{d-3}}{}^{fcd} h_{hf}E_{gc}\eta_d
\label{eq:rel1}
\end{align}
and 
\begin{align}
\mathcal{X}_{abcd}h_{w}^{~a}h_{x}^{~b}h_{y}^{~c}h_{z}^{~d}
=&  
\;
\Omega^{(5-d)/2}
{}^{(d-1)}C_{wxyz} 
\notag \\
&- \frac{2}{d-3}\left(E_{w[y}h_{z]x}-E_{x[y}h_{z]w}\right). \label{C-decomp}
\end{align}
In Eqs.~(\ref{eq:rel1}) and (\ref{C-decomp}), we used the fact that the extrinsic curvature of $\Omega={\rm 
constant}$ surface at $i^0$ is  
\begin{align}
\pi_{ab} & \equiv  (1/2) \mbox \pounds_\eta h_{ab} \nonumber \\
  &=  \frac{1}{2}
(\eta^c \nabla_c h_{ab}+h_{ac}\nabla_b \eta^c+h_{bc} \nabla_a \eta^c )\nonumber \\
 & =   \Omega^{-1/2}h_{ab}\;.
\end{align}
For the derivation of Eq.~(\ref{C-decomp}), see Eq.~(A6) in~\cite{SMS}. (Note that 
the magnetic part defined there is different from ours.)

\subsection{Derivation of Eqs.~(\ref{elepo}) and (\ref{eq:mag2})}
\label{App:secondset}
Hereafter in this appendix, we derive Eqs.~(\ref{elepo}) and (\ref{eq:mag2}).
Firstly,
to facilitate the derivation, we derive the following relation:
\begin{equation}
\mathcal{X}_{abcm}\eta^m =\frac{1}{d-2}\Bigl[
\Omega^{1/2}\nabla_{[b}T_{a]c}+(d-5)\eta_{[b}T_{a]c} \Bigr],
\label{keq}
\end{equation}
where $T_{ab}\equiv \Omega^{(5-d)/2}S_{ab}$ is a tensor which have directional dependent limit at 
$i^0$, and $S_{ab}$ is defined in Eq.~(\ref{eq:S}).
The manipulation of $g^{md} \times$ Eq.~(\ref{eq:a2}) implies 
\begin{equation}
\nabla_m C_{abc}{}^{m}=(d-3)\Omega^{-1}C_{abcp}\nabla^p \Omega\;.
\end{equation}
Note that Eq.~(\ref{eq:a}) was derived from the Bianchi identity 
in the physical vacuum spacetime ($\nabla_{[m}C_{ab]cd}=0$). On the other hand, from 
the Bianchi identity in the unphysical spacetime ($\hat\nabla_{[m}\hat R_{ab]cd}=0$), we can derive 
\begin{equation}
\nabla_m C_{abc}{}^{m}+\frac{2(d-3)}{d-2}\nabla_{[a}S_{b]c}=0~. 
\label{unb}
\end{equation}
From these two equations, we can see that
\begin{equation}
C_{abcm}\eta^m =-\frac{1}{d-2}\Omega^{1/2} \nabla_{[a}S_{b]c} \;.
\end{equation}
It is easy to see that Eq.~(\ref{keq}) holds from this equation. 

Now we are ready 
to derive Eqs.~(\ref{elepo}) and (\ref{eq:mag2}). Let us first take the 
manipulation of $h_p^a h_q^c \eta^b \times$ Eq.~(\ref{keq}), which results in
\begin{align}
E_{pq}& =  \frac{1}{d-2}h_p^ah_q^c\eta^b\Bigl[ 
\Omega^{1/2}\nabla_{[b}T_{a]c}+(d-5)\eta_{[b}T_{a]c}
\Bigr] \nonumber \\
& =   -\frac{1}{2(d-2)}\Bigl[ 
D_p Q_q +h_{pq}E+(4-d)U_{pq} \Bigr] \nonumber \\
& =  -\frac{1}{2(d-2)}
\Bigl[ \frac{1}{d-3}D_p D_q E+h_{pq}E \nonumber \\
& 
\qquad\qquad\qquad\qquad\qquad\qquad\;
+(4-d)U_{pq} \Bigr].
\end{align}
This is Eq.~(\ref{elepo}). 
In the last line, we used the relation derived by the manipulation of 
$h^a_e \eta^b \eta^c \times$ Eq.~(\ref{keq}):
\begin{equation}
D_e E=(d-3)Q_e\;. \label{eq-rela}
\end{equation}
Next, let us apply
$\Omega^{(4-d)/2} \mathcal{E}_{a_1 \cdots a_{d-3}}{}^{fab}\eta_f$ to Eq.~(\ref{keq}).
Then we obtain Eq.~(\ref{eq:mag2}):
\begin{align}
 & B_{a_1 \cdots a_{d-3}c} \nonumber \\
 & ~~=  \frac{1}{d-2}\mathcal{E}_{a_1 \cdots a_{d-3}}{}^{fab}\eta_f 
\Omega^{(4-d)/2} (D_bU_{ac}+h_{bc}Q_a) \nonumber \\
 & ~~=  -\frac{1}{d-2}\mathcal{E}_{a_1 \cdots a_{d-3}}{}^{fab}\eta_f 
\Omega^{(4-d)/2}
D_a \Bigl(U_{bc}-\frac{h_{bc}}{d-3}E \Bigr).
\end{align}

\section{$(d-1)+1$ decomposition}
\label{App:d+1}
In this appendix, 
we show that $d$-momentum defined in Eq.~(\ref{eq:g}) agrees with the ADM 
formulae for energy and momentum:
\begin{align}
E &=  \frac{1}{16\pi G_{d}}\lim_{r_{0}\rightarrow \infty} \int_{ S^{d-2} }
\left( \partial^{a}h_{ab} - \partial_{b}h^{a}_{~a}\right) dS^{b}
\Big|_{r=r_{0}}, 
\\
Q_{N^{a}} &=   \frac{-1}{8\pi G_d} \lim_{r_{0}\rightarrow\infty} 
\int_{S^{d-2}} 
\left(K_{ab}-K^{m}_{\,\,m}h_{ab}\right)N^{a}dS^{b}\Big|_{r=r_{0}}, 
\label{eq:moment}
\end{align}
where $h_{ab}\equiv g_{ab}+t^at^b$ and $K_{ab}\equiv h_{a}^{~c}h_{bd}\nabla_ct^d$ are the induced metric and 
the extrinsic curvature of a $t=\text{constant}$ surface whose unit normal is $t^a$,
and $\partial_a$ is a coordinate derivative with respect to asymptotic Cartesian coordinates.
$N^{a}$ is an asymptotic spacelike translational Killing vector such that $D_{a}N_{b}\rightarrow 0$ 
as $r\rightarrow \infty$, where $D_{a}$ is the connection for $h_{ab}$. 

We also show in this appendix that the angular momentum defined in Eq.~(\ref{eq:h}) transforms in 
translational transformation as Eq.~(\ref{eq:i}).
This appendix may be 
regarded as an extension of the work by Ashtekar and Magnon in four dimensions~\cite{AM}. 
We will describe in much detail because it is very hard to check their result.


\subsection{Energy}
\label{App:energy}
First, let us consider the energy. Let $\hat{\Sigma}$ be a spacelike hypersurface in $\hat{M}$ on $i^{0}$
which has unit timelike vector $\hat{t}^{a}$ as its normal. Then, the energy defined by Eq.~(\ref{eq:g}) becomes  
\begin{equation}
-P_{a}\hat{t}^{a} =- \frac{1}{8\pi G_{d}(d-3)} \int_{S^{d-2}} \hat E_{ab} \hat{t}^{a}\hat{t}^{b} dS\;,
\label{eq:energy}
\end{equation}
where $dS$ is the volume element of a $(d-2)$-dimensional unit sphere $S^{d-2}$. 
In order to compare the above with the ADM formula, 
we must write it down in terms of quantities in physical space-time $M$. 
To do so, we introduce a spacelike hypersurface $\Sigma$ in $M$, 
unit timelike vector $t^{a}$ normal to $\Sigma$, 
and 
a unit radial vector $\eta^{a}=\partial^{a}r$.
$t^a$ and $\eta^a$ are related to 
$\hat{t}^{a}$ and the unit radial vector in the unphysical space-time $\hat\eta^a$ as
$\lim_{r_0\rightarrow \infty}\Omega^{-1}t^{a}=\hat{t}^{a}$
and $\lim_{r_0\rightarrow \infty}\Omega^{-1}\eta^{a}=\hat{\eta}^{a}$, respectively. 
Then, the above expression of energy~(\ref{eq:energy}) becomes 
\begin{align}
-P_{a}\hat{t}^{a} =&  -\frac{1}{8\pi G_{d}(d-3)} 
\lim_{r_{0}\rightarrow \infty}\int_{r=r_{0} }
\!\!\!\!\!\!\!\!
r^{d-1}C_{abcd}\eta^{b}\eta^{d}t^{a}t^{c}dS\,,
\end{align}
where we used the fact that $\Omega \simeq 1/r^2$ near $i^0$. 

Now, we define the usual electric part of the Weyl tensor $e_{ab}\equiv C_{ambn}t^{m}t^{n}$ in 
the physical space-time $M$. This electric part can be decomposed as
\begin{align}
e_{ab} =&\, ^{(d-1)}R_{ab} -K_{a}^{~m}K_{bm}+KK_{ab} \nonumber \\ 
&-\frac{1}{d-2}\Bigl( (d-3)h_{a}^{~m}h_{b}^{~n}+h_{ab}h^{mn}\Bigr) S_{mn}\;.
\end{align}
Taking into account of asymptotic behaviors $K_{ab}=\mathcal{O}( 1/r^{d-2})$ and  
$^{(d-1)}R_{ab}= \mathcal{O} (1/r^{d-1})$ for $r\rightarrow \infty\,$, and the vacuum Einstein equation $R_{ab}=0$, we  
obtain  
\begin{equation}
-P_{a}\hat{t}^{a} = -\frac{1}{8\pi G_{d}(d-3)}
\lim_{r_{0}\rightarrow \infty}
\int_{r=r_{0} }
\!\!\!\!\!\!
r^{d-1} \,^{(d-1)}R_{ab}\eta^{a}\eta^{b}dS. \label{eq:j}
\end{equation}
In order to integrate by parts in direction $r$, 
we rewrite the integral into the following form:
\begin{align}
-P_{a}\hat{t}^{a} =& -\frac{1}{8\pi G_{d}(d-3)}\lim_{r_{0}\rightarrow \infty} \label{eq:k} \\ 
&\times  \frac{1}{\Delta r}\int_{r=r_{0}}^{r=r_{0}+\Delta r}\int_{S^{d-2}}r^{d-1} \,^{(d-1)}R_{ab}\eta^{a}\eta^{b}drdS, \notag
\end{align}
where we used the fact that the integrand in Eq.~(\ref{eq:j}) is independent of $r$ at large $r$.
In this expression, the part which contribute to the integral is
\begin{equation}
^{(d-1)}R_{ab} \sim \frac{1}{2}
\left( 
\partial^c \partial_b h_{ac} +\partial_a \partial^c h_{bc} - \partial^c \partial_c h_{ab} - \partial_a \partial_b h^c_{~c}
\right).
\label{eq:l}
\end{equation}
Substituting (\ref{eq:l}) into (\ref{eq:k}) and integrating by parts, we can get the desired result. 
Since this calculation is a little difficult, 
we describe carefully. First, we integrate the first part in (\ref{eq:l}) by parts:
\begin{align}
&\frac{1}{\Delta r}
\!\!
\int_{S^{d-2}\times \Delta r}
r
\left(
\partial^c\partial_b h_{ac}
\right)
\eta^{a}\eta^{b}dV
\label{eq:first} 
\\ 
=&\frac{1}{\Delta r}
\!\!
\int_{S^{d-2}\times \Delta r}
\!\Bigl[
\partial^c\bigl(r\left(\partial_bh_{ac}\right)
\eta^{a}\eta^{b}\bigr)
\!-\!
\left(\partial_bh_{ac}\right) \partial^c(r\eta^{a}\eta^{b})
\Bigr] dV,
\notag
\end{align}
where $dV\equiv r^{d-2}drdS$. The first term in the right-hand side becomes 
\begin{align}
\frac{1}{\Delta r}&\int_{S^{d-2}\times \Delta r}
\partial^c\bigl(r\left(\partial_bh_{ac}\right)\eta^{a}\eta^{b}\bigr)dV 
\notag \\
&=\;\;\;\,
\frac{1}{\Delta r}\int_{S^{d-2}}
r\left(\partial_bh_{ac}\right)
\eta^{a}\eta^{b}dS^{c}   \Big|_{r=r_0+\Delta r}
\notag \\
&\;\;\;\;
-\frac{1}{\Delta r}\int_{S^{d-2}}r
\left(\partial_b h_{ac}\right)
\eta^{a}\eta^{b}dS^{c} \Big|_{r=r_0}
\notag \\ 
&= \int_{S^{d-2}}
\left(\partial_b h_{ac}\right)
\eta^{a}\eta^{b}dS^{c} \Big|_{r=r_0} \quad ,
\end{align}
where $dS^{c}\equiv \eta^{c}r^{d-2}dS$. In the first and the second equalities, we 
used the Gauss theorem, and the fact that $\left(\partial_b h_{ac}\right)\eta^{a}\eta^{b} r^{d-2}$ 
is independent of $r$ in the limit of $r_0 \to \infty$. 
The second term in the right-hand side of Eq.~(\ref{eq:first}) becomes
\begin{align}
 & \frac{1}{\Delta r}\int_{S^{d-2}\times \Delta r}
\left(\partial_b h_{ac}\right)
\partial^c(r\eta^a \eta^b)dV \nonumber \\
 & = \frac{1}{\Delta r}\int_{S^{d-2}\times \Delta r}
\!\!\!\!\!
\left(\partial_b h_{ac}\right)
(\eta^a \eta^b \eta^c+q^{ac}\eta^b+q^{bc}\eta^a)r^{d-2}drdS \nonumber \\
 & = \int_{S^{d-2}}
\!\!
\left(\partial_b h_{ac}\right)
(\eta^a \eta^b \eta^c+q^{ac}\eta^b+q^{bc}\eta^a)r^{d-2}dS.
\end{align}
To transform the second into the third line, we used the fact that the integrand 
in the second line does not depend on $r$. Then, we obtain 
\begin{align}
&\int_{S^{d-2}}r^{d-1}
\left( \partial^c\partial_b h_{ac} \right)
\eta^{a}\eta^{b}dS \nonumber \\
&\;\;
=-\int_{S^{d-2}}
\left(\partial_b h_{ac}\right)
(\eta^{a}q^{bc}+\eta^{b}q^{ac})r^{d-2}dS.
\end{align}
Here, we defined a metric $q_{ab}$ on $r=\text{constant}$ surface such that
$\partial_{a}\eta_{b}=(h_{ab}-\eta_{a}\eta_{b})/r\equiv q_{ab}/r$. In the 
same way, the other terms in (\ref{eq:l}) are transformed as 
\begin{align}
 & \int_{S^{d-2}} r^{d-1}
\left( \partial_a\partial^c h_{bc} \right)
\eta^a \eta^b dS
=  -(d-2)\int_{S^{d-2}} \partial^c h_{ac}dS^a, \nonumber \\
 & \int_{S^{d-2}} r^{d-1}
\left( \partial^c \partial_c h_{ab} \right) \eta^a \eta^b dS \nonumber \\
& \qquad\qquad\qquad
=  -\int_{S^{d-2}}
\left( \partial_c h_{ab} \right)
(q^{ac}\eta^b+q^{bc}\eta^a)r^{d-2}dS\;, \nonumber \\
 & \int_{S^{d-2}} r^{d-1}
\left(
\partial_a\partial_b h^c_{~c}
\right)
\eta^a \eta^b dS
=  -(d-2)\int_{S^{d-2}}
\!
\partial_a h^c_{~c}
dS^a.\nonumber 
\end{align}
Finally, we obtain the desired result: 
\begin{equation}
-P_{a}\hat{t}^{a} = \frac{1}{16\pi G_{d}}\lim_{r_{0}\rightarrow \infty} \int_{r=r_{0}}\left( \partial ^{a}h_{ab} - \partial_{b}h^{a}_{~a}\right) dS^{b} \; .
\end{equation}

\subsection{Momentum}
\label{App:momentum}
Next, let us consider momentum. The components of $(d-1)$-momentum along a spacelike vector $N^{a}$ at $i^{0}$ can be written as
\begin{equation}
P_{a}\hat{N}^{a} = \frac{1}{8\pi G_{d}(d-3)} \int_{S^{d-2}}\hat E_{ab}\hat{N}^{a}\hat{t}^{b}dS. 
\end{equation}
In terms of quantities of physical space-time, this equation becomes 
\begin{equation}
P_{a}\hat N^{a}=\frac{1}{8\pi G_{d}(d-3)}\lim_{r\rightarrow \infty} \int_{r=r_{0}}r^{d-1}C_{abcd}\eta^{b}\eta^{d}N^{a}t^{c}dS\,,
\end{equation}
where $N^{a}=\lim_{\rightarrow i^{0}}\Omega\hat N^{a}$. 
Using the Codacci equation and the vacuum Einstein equation, this expression becomes 
\begin{align}
P_{a}\hat{N}^{a}=\;&\frac{1}{8\pi G_{d}(d-3)}\lim_{r_{0}\rightarrow \infty} \int_{r=r_{0}}r^{d-1} 
 \notag \\ 
&\times \left(D_{d}K_{ab}-D_{a}K_{db}\right) \eta^{b}\eta^{d}N^{a}dS. 
\label{p-formula}
\end{align}
Using the fact that the leading part of $r^{d-1}D_dK_{ab}$ does not depend on 
$r$, the first term in the right-hand side is reexpressed as volume 
integral as 
\begin{align}
 & \int_{r=r_0}r^{d-1}\left(D_dK_{ab}\right)
\eta^b \eta^d N^a dS \nonumber \\
  &\;\;
=\frac{1}{\Delta r} \int_{S^{d-2}\times \Delta r} 
 r \left( D_d K_{ab} \right)
\eta^b \eta^d N^a dV \nonumber \\
 &\;\;
=\frac{1}{\Delta r} 
\int_{S^{d-2}\times \Delta r} \Bigl[ 
\;
D_d(rK_{ab}\eta^b \eta^d N^a) \nonumber \\
 & \;\;\qquad\qquad\qquad\quad\;\;
-K_{ab}D_d(r\eta^b \eta^d N^a) \;\Bigr]dV.
\label{p-last}
\end{align}
Using the Gauss theorem to the first term in the last line, we see that 
\begin{align}
&\frac{1}{\Delta r} 
\int_{S^{d-2}\times \Delta r} 
D_d(rK_{ab}\eta^b \eta^d N^a) 
dV
\notag \\
 &=\frac{1}{\Delta r}\biggl[
\int_{S^{d-2}}
\!\!\!\!\!
rK_{ab}N^a dS^b
\Big|_{r=r_0+\Delta r}
\!-\!
\int_{S^{d-2}}
\!\!\!\!\!
rK_{ab}N^a dS^b 
\Big|_{r=r_0}
\biggr] \nonumber \\
 & =\int_{S^{d-2}}K_{ab}N^a dS^b
\Big|_{r=r_0}
\;\;,
\end{align}
where we used the fact that $K_{ab}N^a r^{d-2}$ does not depend on $r$. 
The second term in Eq.~(\ref{p-last}) can be rearranged as 
\begin{align}
  -&\frac{1}{\Delta r}
\int_{S^{d-2}\times \Delta r}K_{ab}D_d(r\eta^b \eta^d N^a)dV
\nonumber \\
  &=-\frac{d-1}{\Delta r}\int_{S^{d-2}\times \Delta r}K_{ab}\eta^bN^adV \nonumber \\
  &=-(d-1)\int_{S^{d-2}}K_{ab}N^adS^b \Big|_{r=r_0}
\;\;. 
\end{align}
In the same way, the second term of Eq.~(\ref{p-formula}) is
rearranged as 
\begin{align}
 & \int_{r=r_0} r^{d-1} \left( D_a K_{db} \right)
\eta^b \eta^d N^a dS \nonumber \\
 & = 2\int_{S^{d-2}} K_{ab}\eta^a \eta^b N^c dS_c -2\int_{S^{d-2}} K_{ab}N^a dS^b
\label{p-m}
\end{align}
From this equation, we can obtain a relation
\begin{align}
2&\lim_{r_{0}\rightarrow \infty} \int_{r=r_{0}}r^{d-2}K_{ab}N_{c}\eta^{a}\eta^{b}\eta^{c}dS \notag \\ 
=&\lim_{r_{0}\rightarrow \infty} \int_{r=r_{0}}\Big( K_{ab}-(d-3)Kh_{ab}\Big)N^{a}dS^{b}.
\label{K-relation}
\end{align}
Derivation of this relation is a little non-trivial, so we describe it in detail. 
Note that the Gauss theorem makes the surface integral into the volume integral as 
\begin{align}
 & \int_{r=r_{0}}r^{d-2}K_{ab}N_{c}\eta^{a}\eta^{b}\eta^{c}dS \nonumber \\ 
 & \!
=\!\frac{1}{\Delta r}
\!
\int_{S^{d-2}\times \Delta r} \partial^a
\left( rK_{ab}N_{c}\eta^{b}\eta^{c}\right)dV \nonumber \\
 & \!
=\!\frac{1}{\Delta r}
\!
\int_{S^{d-2}\times \Delta r}
\Bigl[ r \left(D_bK \right) \eta^b N^c\eta_c 
\! + \!
K_{ab}D^a\left(r\eta^b N^c \eta_c\right) \Bigr]dV, \label{K-relation2}
\end{align}
where we used the momentum constraint equation for the vacuum 
Einstein equation, $D_aK^{a}_b-D_bK=0$, in the last line. 
The first and the second terms are rearranged respectively as
\begin{align}
 \frac{1}{\Delta r}
&\int_{S^{d-2}\times \Delta r}r \left(D_bK \right)
\eta^b N^c\eta_c dV
\nonumber \\
  = &\frac{1}{\Delta r}\int_{S^{d-2}\times \Delta r}
\Bigl[D_b\left(rK\eta^b N^c \eta_c\right)-KD_b\left(r\eta^b N^c \eta_c\right)  \Bigr]dV \nonumber \\
 =& -(d-2)\int_{S^{d-2}}KN^a dS_a \Big|_{r=r_0}\;\; ,
\end{align}
and
\begin{align}
 & \frac{1}{\Delta r}\int_{S^{d-2}\times \Delta r} K_{ab}D^a\left(r \eta^b N^c \eta_c\right)dV
\nonumber \\
 & \!= \!
\int_{S^{d-2}}
\!\!
\left(KN^adS_a+K_{ab}N^bdS^a -K_{ab}\eta^a \eta^b N^cdS_c\right)\Big|_{r=r_0}
.
\end{align}
Then, we proceed as 
\begin{align}
 & \int_{r=r_{0}}
r^{d-2}
K_{ab}N_{c}\eta^{a}\eta^{b}\eta^{c}dS \nonumber \\ 
 & ~~=~
\int_{S^{d-2}} 
\Bigl(
K_{ab}N^b-(d-3)KN_a 
\Bigr)
dS^a\Big|_{r=r_0}
\nonumber \\
 & ~~~~ -\int_{S^{d-2}}K_{ab}\eta^a \eta^b N^c dS_c
\Big|_{r=r_0}
\;\;. 
\end{align}
The last term in the right-hand side is the same with the left-hand side 
except for the signature. Therefore, we have the relation of Eq.~(\ref{K-relation}). 

Substituting Eq.~(\ref{K-relation}) into Eq.~(\ref{p-m}), we can show 
\begin{align}
 & \int_{r=r_0}r^{d-1}(D_dK_{ab}-D_aK_{db})\eta^b \eta^d N^a dS \nonumber \\
 & =-(d-3)\int_{S^{d-2}}(K_{ab}-Kh_{ab})N^a dS^b \Big|_{r=r_0}\;\;,
\end{align}
and then, combining this equation with Eq.~(\ref{p-formula}), we see that
 our formula~(\ref{eq:g}) for $(d-1)$-momentum becomes the ADM formula, that is
\begin{equation}
P_{a}\hat{N}^{a} =- \frac{1}{8\pi G_{d}}\lim_{r_{0}\rightarrow \infty} \int_{r=r_{0}}\left( K_{ab}-Kh_{ab}\right)N^{a}dS^{b}.
\end{equation}

\subsection{Angular momentum}
\label{App:angular}
Finally, we consider translational transformation of angular momentum $M_{ab}$. 
We consider translation $\omega_{a}$ which is a 
fixed vector at $i^{0}$, and relate it with $\alpha$ as $\alpha=\omega_{a}\hat{\eta}^{a}$. This translation transforms $\hat{\eta^{a}}$ as
\begin{equation}
\hat{\eta}_{a}'
\;\hat{=}\;
\hat{\eta}_{a} + \frac{1}{2}\Omega^{(d-3)/2}\left((d-2)\alpha\hat{\eta}_{a}+\Omega^{1/2}\hat{\nabla}_{a}\alpha\right)\; .
\end{equation}
Then, the magnetic part of the Weyl tensor $\hat \beta_{a_{1}\cdots a_{d-3}b}$ transforms as
\begin{align}
\hat \beta_{a_{1}\cdots a_{d-3}b}' \,\hat=& \; \hat \beta_{a_{1}\cdots a_{d-3}b} 
+\frac{d-2}{d-3}\hat \epsilon_{a_{1}\cdots a_{d-3}mpq}\hat\eta^m\hat E^{q}_{~b}\hat D^{p}\alpha 
\notag \\
&+ 
\frac{1}{d-3}\hat \epsilon _{a_{1}\cdots a_{d-3}mpb}\hat\eta^m\hat E^{pr}\hat D_{r}\alpha \; . \label{eq:beta}
\end{align}
We used the projection formulae of the Weyl tensor~(\ref{eq:rel1}) and (\ref{C-decomp}) to derive this equation.
Substituting (\ref{eq:beta}) into (\ref{eq:h}), 
and noting that 
$F_{ab} \hat \epsilon^{b}_{~e_{1}\cdots e_{d-2}m}\hat\eta^m dS^{e_{1}\cdots e_{d-2}}$ vanishes
since $d-1$ indices of $\hat\epsilon$ are projected onto $(d-2)$-dimensional surface,
we find that usual translational transformation
\begin{equation}
M_{ab}'=M_{ab}+2P_{\,[a}\hat \omega_{b]}\; ,
\end{equation}
where $\hat D_{a}\alpha=\hat \omega_{a}$, is correctly reproduced including the coefficient,
if we define the angular momentum as (\ref{eq:h}).

\vfill


\end{document}